\documentclass[12pt,a4paper]{article}
\usepackage{amsfonts,latexsym}
\usepackage{amsmath,amssymb}
\usepackage{graphicx,color}

\renewcommand{\thefootnote}{\fnsymbol{footnote}}

\begin{document}

\begin{flushright}
CTPU-PTC-26-05
\end{flushright}

\vspace{12mm}
\begin{center}
{{{\Large {\bf Gauss-Bonnet  scalarization of charged qOS-black holes }}}}\\[10mm]
{Hong Guo$^1$\footnote{e-mail address: guohong@ibs.re.kr}, Wontae Kim$^{2,3}$\footnote{e-mail address: wtkim@sogang.ac.kr}, and Yun Soo Myung$^{3}$\footnote{e-mail address: ysmyung@inje.ac.kr}}\\[8mm]
{${}^1$Particle Theory and Cosmology Group, Center for Theoretical Physics of the Universe, \\
 Institute for Basic Science (IBS), Daejeon 34126, Republic of Korea\\[0pt]}
{${}^2$Department of Physics, Sogang University, Seoul, 04107, Republic of Korea \\[0pt]}
{${}^3${Center for Quantum Spacetime, Sogang University, Seoul 04107, Republic of  Korea\\[0pt] }}

\vspace{12mm}

\end{center}
 \begin{abstract}
    \noindent The Gauss-Bonnet (GB)  scalarization  for  charged quantum Oppenheimer-Snyder (cqOS)-black holes is investigated  in the Einstein-Gauss-Bonnet-scalar theory with  the nonlinear electrodynamics (NED) term.
    Here, the scalar coupling function to GB term is given by $f(\phi)=2\lambda \phi^2$ with a coupling constant $\lambda$.
    Three parameters of
    mass ($M$), action parameter ($\alpha$), and  magnetic charge ($P$) are necessary to describe the cqOS-black hole, and it may  become  the  qOS-black hole when $P=M$.
    The GB scalarization of cqOS-black holes comes into two cases GB$^\pm$, depending on the sign of GB term which triggers the different phenomena.
    For $\alpha=0$ and $\lambda>0$, GB$^+$ scalarization is allowed, while for $\alpha\not=0$ and $\lambda<0$,  GB$^-$ scalarization appears for a narrow band of $3.5653\le \alpha\le 4.6875$.
    After discussing the onset GB$^-$ scalarization,  we construct scalarized cqOS-black holes which belong to the single branch. 
    The scalar field decays much more rapidly compared to the GB$^+$ case.
    Stability analysis shows these scalarized black holes are linearly stable under scalar perturbations.
\end{abstract}


\newpage
\renewcommand{\thefootnote}{\arabic{footnote}}
\setcounter{footnote}{0}

\vspace{2mm}


\section{Introduction}
\label{sec: Intro}
It is well-known as no-hair theorem that a black hole can be described  by  three classical parameters of  mass, electric charge, and angular momentum in Einstein gravity
\cite{Carter:1971zc,Ruffini:1971bza}. Considering a minimally coupled scalar field to  gravitational and Maxwell  fields with a positive scalar potential, there are no black hole solutions with scalar hair~\cite{Herdeiro:2015waa}. However, its evasion  appeared immediately  for  a nonminimally coupled scalar  to either Gauss-Bonnet (GB) term~\cite{Doneva:2017bvd,Silva:2017uqg,Antoniou:2017acq}, or Maxwell (M) term  with a positive (+) coupling constant~\cite{Herdeiro:2018wub,Myung:2018vug,Myung:2018jvi}.  The former case denotes curvature-induced (GB$^+$) spontaneous scalarization of Schwarzschild black hole, while the latter represents  charged-induced (M$^+$) spontaneous scalarization of Reissner-Nordtr\"om black hole.  In these cases  triggered   by tachyonic instability, one obtained infinite branches of scalarized  black holes from infinite scalar clouds.

For the negative ($-$) coupling constant, the  spin-induced (GB$^-$) scalarization
of Kerr black holes with mass $M$ and rotation parameter $a$  were studied  in the Einstein-Gauss-Bonnet-Scalar (EGBS)
theory~\cite{Cunha:2019dwb,Collodel:2019kkx}.
Here, an $a$-bound of $ a_c(=0.5)\le a \le1$ with $M=1$ was found to describe  the condition for  onset of GB$^-$ scalarization~\cite{Dima:2020yac,Hod:2020jjy,Zhang:2020pko,Doneva:2020nbb}.  This indicates that  GB$^-$ scalarization is forbidden  for the low rotation of $0<a<a_c$.
Furthermore, the condition for spin-induced (M$^-$) scalarization of Kerr-Newman black holes was found to be  $ a_c(=0.672)\le a\le 1$ for mass $M=1$ and  charge $Q=0.4$~\cite{Lai:2022spn} as well as  the spin-charge  induced scalarization of Kerr-Newman black holes was studied in Einstein-Maxwell-Scalar (EMS) theory~\cite{Lai:2022ppn}.
 Also, it is  notified that  the dual coupling concept of curvature and matter-induced scalarization discussed in the EGBMS theory~\cite{Belkhadria:2025lev}.

Recently, the quantum Oppenheimer-Snyder (qOS) black hole was found by studying the qOS gravitational collapse in conjunction with the loop quantum cosmology~\cite{Lewandowski:2022zce}, providing a viable, singularity-free scenario that modifies Schwarzschild geometry while remaining consistent with it at large distances.
It is used as a testbed to study quantum-gravity effects on horizons~\cite{Piechocki:2020bfo,Shi:2024vki,Dong:2024hod}, radiation~\cite{Gong:2023ghh,Zinhailo:2024kbq}, stability~\cite{Ou:2025bbv}, shadows~\cite{Ye:2023qks,Luo:2024nul}, and gravitational waves~\cite{Yang:2025esa}, bridging the gap between loop quantum gravity and potentially observable astrophysical signatures.
An undesirable point to this approach is that  an  explicit action  $\mathcal{L}_{\rm qOS}$ is not known and thus,   its energy momentum tensor was used to obtain qOS black hole.
Explicitly, the qOS-black hole was obtained by applying the junction conditions but not solving the equations of motion.
In addition, quantum parameter-mass induced (GB$^-$) scalarization was studied in the EGBS theory~\cite{Chen:2025wze, Myung:2025pmx}, showing a deep connection  between thermodynamics of qOS-black hole and onset GB$^-$ scalarization.

However, one could not make a further progress to obtain its scalarized qOS-black holes  because of unknowing its action $\mathcal{L}_{\rm qOS}$.
Regarding to  appropriate actions for $\mathcal{L}_{\rm qOS}$,  a candidate was proposed  by considering  a nonlinear electrodynamics (NED) term. 
In this case, one found cqOS-black hole solution and  the qOS black hole solution may be  recovered when imposing  $P=M$ which is not the extremal condition~\cite{Mazharimousavi:2025lld}.
Inspired by this work, we will investigate scalarized  cqOS-black holes in the EGBS-NED theory with a scalar coupling $f(\phi)$ to the GB term.

The organization of the present work is as follows.
In section 2, we obtain  the cqOS- black holes. We briefly mention  thermodynamics of the cqOS-black hole described by mass ($M$) and action parameter ($\alpha$) with a fixed  magnetic charge $P$.
Section 3 is devoted to discussing  onset  GB$^-$ scalarization of cqOS-black holes.
Here, one finds that two Davies quantities of heat capacity are not identified with   two critical onset  parameters  for GB$^-$ scalarization.
This implies that this  black hole is not really   considered as  the qOS black hole.
We will obtain  scalarized cqOS-black holes through GB$^-$ scalarizations in section 4.
Section 5 is focused on performing the stability analysis of scalarized cqOS-black holes.


\section{cqOS-black holes and its thermodynamics}
The EGBS-NED theory is introduced as 
\begin{equation}
\mathcal{L}_{\rm EGBS-NED}=\frac{1}{16 \pi}\Big[ R-2\partial_\mu \phi \partial^\mu \phi+ f(\phi) {\cal R}^2_{\rm GB}+ {\cal L}_{\rm NED}\Big]\label{Action1}
\end{equation}
where ${\cal R}^2_{\rm GB}$ denotes the GB term of $R^2-4R_{\mu\nu}R^{\mu\nu}+R_{\mu\nu\rho\sigma}R^{\mu\nu\rho\sigma}$
and  the NED term is given by
$\mathcal{L}_{\rm NED}=-2\xi(\mathcal{F})^{\frac{3}{2}}$
with $\mathcal{F}=F_{\mu\nu}F^{\mu\nu}$.
To explore  GB$^\pm$ scalarizations, a quadratic scalar coupling $f(\phi)= 2\lambda \phi^2$ is chosen with  positive/negative coupling constant $\lambda$.

We obtain the Einstein equation from the EGBS-NED theory
\begin{eqnarray}
 G_{\mu\nu}=2\partial _\mu \phi\partial _\nu \phi -(\partial \phi)^2g_{\mu\nu}+\Gamma_{\mu\nu}+T^{\rm NED}_{\mu\nu}, \label{equa1}
\end{eqnarray}
where the GB-scalar coupling term  $\Gamma_{\mu\nu}$   takes the form with $\Psi_{\mu}= f'(\phi)\partial_\mu \phi$
\begin{eqnarray}
\Gamma_{\mu\nu}&=&2R\nabla_{(\mu} \Psi_{\nu)}+4\nabla^\alpha \Psi_\alpha G_{\mu\nu}- 8R_{(\mu|\alpha|}\nabla^\alpha \Psi_{\nu)} \nonumber \\
&+&4 R^{\alpha\beta}\nabla_\alpha\Psi_\beta g_{\mu\nu}
-4R^{\beta}_{~\mu\alpha\nu}\nabla^\alpha\Psi_\beta.  \label{equa2}
\end{eqnarray}
In addition, the  NED energy-momentum tensor is given by
\begin{equation}
T^{\rm NED}_{\mu\nu}=6\xi \sqrt{\mathcal{F}}\Big[F_{\mu\lambda}F_\nu^\lambda-\frac{1}{6}\mathcal{F}g_{\mu\nu}\Big]. \label{EM-t}
\end{equation}
The nonlinear-Maxwell equation is given by $\nabla_\mu(\sqrt{\mathcal{F}} F^{\mu\nu})=0$~\cite{Mazharimousavi:2025lld}.
Taking into account  a magnetic field strength $F_{\theta\varphi}=P \sin \theta~(\mathcal{F}=2P^2/r^4)$ and  $\xi=2^{-3/2} 3\alpha /P$,  Eq.(\ref{EM-t})  is given   by an anisotropic  energy-momentum tensor
\begin{equation}
T^{\rm NED,\nu}_\mu=\frac{3\alpha P^2}{r^6}{\rm diag}[-1,-1,2,2]\equiv{\rm diag}[-\rho,p_r,p_\theta,p_\phi].
\end{equation}
It satisfies three (null, weak, strong) energy conditions for $\alpha>0$, except that the dominant energy condition of $\rho\ge |p_i|$ is violated.

On the other hand, the scalar equation indicates
\begin{equation}
\square \phi +\frac{1}{4}f'(\phi) {\cal R}^2_{\rm GB}=0 \label{s-equa}.
\end{equation}
Solving the Einstein equation (\ref{equa1})  without scalar ($\phi=0$), one finds the cqOS-black hole  solution
\begin{equation} \label{ansatz}
ds^2_{\rm cqOS}= \bar{g}_{\mu\nu}dx^\mu dx^\nu=-g(r)dt^2+\frac{dr^2}{g(r)}+r^2d\Omega^2_2
\end{equation}
whose metric function is given by~\cite{Lewandowski:2022zce}
\begin{equation}
g(r)\equiv 1-\frac{2\tilde{m}}{r}=1-\frac{2M}{r}+\frac{\alpha P^2}{r^4}, \label{g-sol}
\end{equation}
where mass function $\tilde{m}$ is defined. From $\tilde{m}=0$, one may define the bouncing radius
\begin{equation}
r_b(M,\alpha,P)=\Big(\frac{\alpha P^2}{2M}\Big)^{1/3}
\end{equation}
which can be  also obtained from the junction condition.
Choosing  $P=M$ leads to the qOS black hole~\cite{Lewandowski:2022zce}, but this fixing seems  to be  unnatural. This choice has nothing to do with the condition for the extremality.
Instead, it  suggests  the nature of the present model as charged quantum Oppenheimer-Snyder model~\cite{Mazharimousavi:2025lld}  because its ($P$) role is similar to $\alpha$ and all expressions pertaining $P$ came out as  ``$\alpha P^2$".
It is interesting to note that the electrically charged  black hole solution was found  for $\mathcal{F}^n$ in the Einstein-NED theory~\cite{Hassaine:2008pw}.

Let us mention briefly  the thermodynamics of cqOS-black hole with Smarr formula by looking for the outer horizon $r_+$.
\begin{figure}
\centering
\mbox{
(a)
\includegraphics[angle =0,scale=0.43]{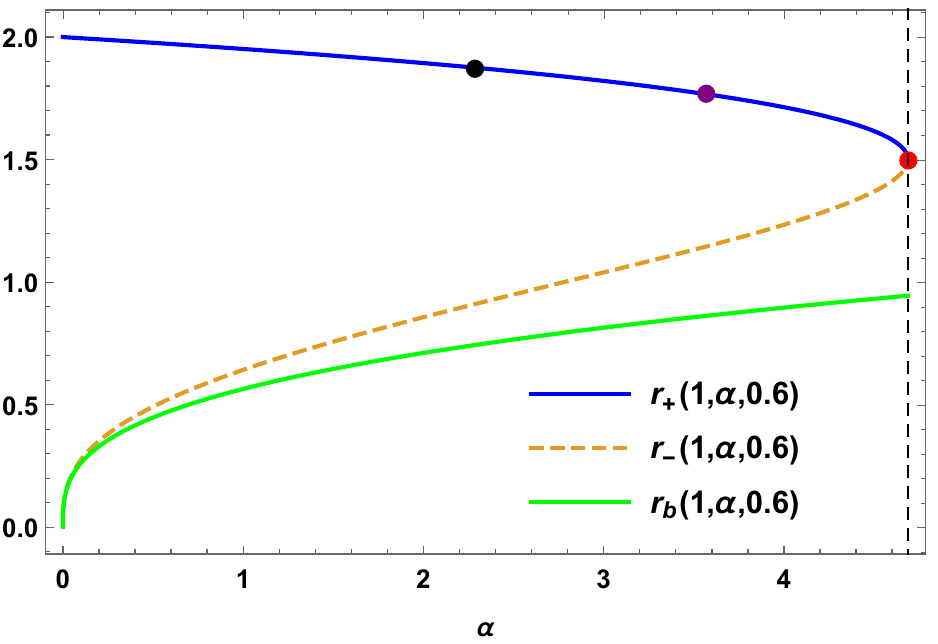}
(b)
\includegraphics[angle =0,scale=0.43]{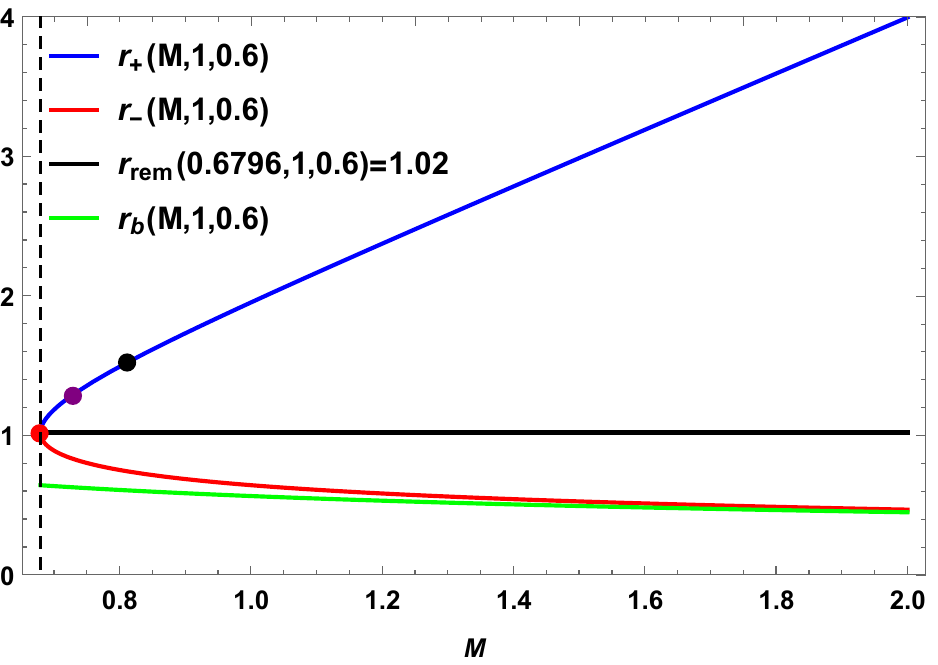}
}
\caption{ (a) Two outer/inner horizons $r_\pm(M=1,\alpha,P=0.6)$ are  functions of $\alpha\in[0,4.6875]$.  The bouncing radius  $r_{b}(1,\alpha,0.6)$ as function of $\alpha$ is inside the inner horizon. Here, $r_+(1,\alpha,0.6)$  involves the Davies point at [(2.29,1.88), black dot], critical onset point [(3.5653,1.7676), purple dot], and extremal point [(4.6875,1.5), red dot]. (b) Two  horizons  $r_{\pm}(M,\alpha=1,P=0.6)$ are   functions of   $M\in[M_{\rm rem}=0.6796,\infty]$, showing the lower mass bound [remnant point at (0.6796,1.02), red dot]. The black dot represents the Davies point at (0.81,1.52), while the purple dot denotes the critical onset point at (0.7277,1.286). The bouncing radius is located inside the inner horizon.}
\label{fig1}
\end{figure}
Solving $g(r)=0$, one finds two real and complex  solutions
\begin{eqnarray}
&&r_i(M,\alpha,P),~{\rm for}~i=4,3,2,1. \label{f-roots}
\end{eqnarray}
Here, we note that $r_4$ is replaced by  $r_+$ and $r_3$ is denoted by  $ r_-$, but  $r_{2/1}$ become complex solutions.
The extremal black hole and the remnant  for $r_+=r_-$ occur for
\begin{equation}
\alpha_e(M,P)=\frac{27 M^4}{16 P^2},\quad M_{\rm rem}(\alpha,P)=2\sqrt{\frac{P\sqrt{\alpha}}{3\sqrt{3}}},
\end{equation}
which leads to $\alpha_e=4.6875$ for $M=1,P=0.6$ and $M_{\rm rem}=0.6796$ for $\alpha=1,P=0.6$.
As is shown in Fig.~\ref{fig1}(a), there is an upper bound  on $\alpha$ (extremal point $\alpha=\alpha_e$  for $M=1$), while there is a lower bound for the mass of  black hole (remnant mass $M=M_{\rm rem}$ for $\alpha=1$) but there is no  upper bound on the mass [see Fig.~\ref{fig1}(b)].

Let us  introduce thermodynamics of cqOS-black hole to make its connection to onset GB$^-$ scalarization.
We find that black hole mass $m=(\alpha P^2+r_+^4)/2r_+^3$, the area-law entropy $S=\pi r_+^2$, the Hawking temperature $T$ and  the heat capacity $C$ with   chemical potential $W_{\alpha}=P^2/r_+^3$ as
\begin{eqnarray}
T(M,\alpha,P)&=&\frac{r_+^4(M,\alpha,P)-3\alpha P^2}{4\pi r_+^5(M,\alpha,P)},  \label{ther2} \\
C(M,\alpha,P)&=-&\frac{2\pi r_+^2(M,\alpha,P)[r_+^4(M,\alpha,P)- 3\alpha P^2]}{ r_+^4(M,\alpha,P)-15 \alpha P^2}.  \label{ther3} 
\end{eqnarray}
\begin{figure}
\centering
\mbox{
(a)
\includegraphics[angle =0,scale=0.43]{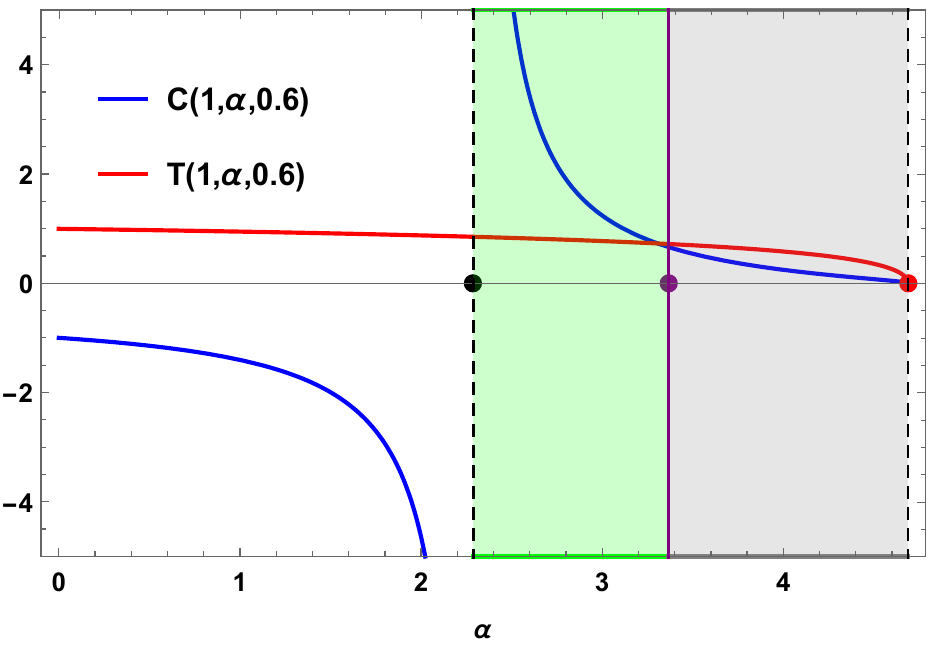}
(b)
\includegraphics[angle =0,scale=0.43]{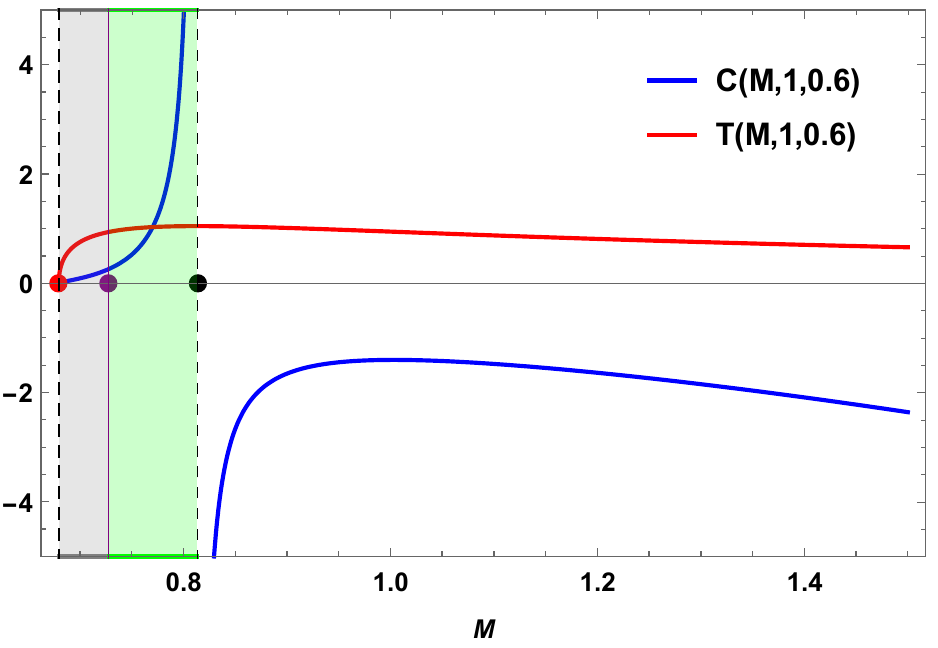}
}
\caption{ Heat capacity $C(M,\alpha,P)/|C_S(1,0,0)|$ with $|C_S(1,0,0)|=25.13$ and  temperature $T(M,\alpha,P)/0.04$.  (a) Heat capacity $C(M=1,\alpha,P=0.6)$ blows up at Davies point ($\alpha_D=2.29,~\bullet$) and it is zero at the extremal point ($\alpha_e=4.6875,$ red dot). The shaded region represents $C\ge 0$ and it is divided  at a line ($\alpha=\alpha_c=3.5653$). (b) Heat capacity $C(M,1,0.6)$ blows up at Davies point ($M_D=0.81,\bullet$). The heat capacity and temperature are zero at the remnant point ($M_{\rm rem}=0.6796,$ red dot). The shaded region denotes  $C\ge 0$, but it is divided at a line ($M=M_c=0.7277$). }
\label{fig2}
\end{figure}
The first-law of thermodynamics  and Smarr formula are satisfied  as
\begin{equation}
dm=TdS+W_\alpha d\alpha,\quad m=2TS+4W_\alpha \alpha,
\end{equation}
which indicates  that the action parameter $\alpha$ plays the role of a thermodynamic variable, instead of  magnetic charge $P$.  Hereafter, we fix $P=0.6$ for convenience.
Considering  Eqs.~(\ref{ther2}) and (\ref{ther3}) together with Fig.~\ref{fig2}, $T$ and $C$ are zero when their numerators are zero, showing extremal and remnant points,  while $C$ blows up when its denominator is zero, leading to the Davies point for a sharp phase transition.  We note that two shaded regions represent positive heat capacity ($C\ge 0$). Each shaded region is divided  at the line ($\alpha=\alpha_c=3.5653/M=M_c=0.7277$) and two gray regions are consistent with the allowed regions for onset GB$^-$ scalarization.

 In this case, we define  the numerator-zero condition  [$nc=0$] and the condition for denominator-zero [$dc=0$] of heat capacity  $C$ defined as
\begin{eqnarray}
 nc(M,\alpha,P)&\equiv&r_+^4(M,\alpha,P)- 3\alpha P^2=0,  \label{rem-c} \\
 dc(M,\alpha,P)&\equiv&r_+^4(M,\alpha,P)- 15\alpha P^2=0.  \label{d-c}
\end{eqnarray}
The former contains the extremal/remnant points ($C\to 0$, termination), while the latter includes the Davies point ($C\to \infty$, a sharp phase transition).

\section{Onset of GB$^-$ scalarization}

In this section, we wish to study the onset GB$^-$scalarization of  cqOS-black hole by computing critical onset mass $M_c$  and action parameter $\alpha_c$.
For this purpose, we start with the linearized scalar  equation:  $(\bar{\square}+ \lambda \bar{{\cal R}}^2_{\rm GB})\delta \phi=0$.
Let us introduce a tortoise coordinate defined by $dr_*=dr/g(r)$ and a separation of variables
\begin{equation}
\delta\phi(t,r_*,\theta,\varphi)=\sum_m\sum^\infty_{l=|m|}\frac{\psi_{lm}(t,r_*)}{r}Y_{lm}(\theta,\varphi).
\end{equation}
The linearized $s(l=0,m=0)$-scalar equation reduces  to
\begin{equation} \label{mode-d}
\frac{\partial^2\psi_{00}(t,r_*)}{\partial r_*^2} -\frac{\partial^2\psi_{00}(t,r_*)}{\partial t^2}=V(r)\psi_{00}(t,r_*),
\end{equation}
where the scalar potential $V(r)$ is given by
\begin{equation} \label{pot-c}
V(r)=g(r)\Big[\frac{2M}{r^3}-\frac{4\alpha P^2}{r^6}+m^2_{\rm eff}(r)\Big]
\end{equation}
with its effective mass term
\begin{equation}
m^2_{\rm eff}(r)\equiv -\lambda \bar{\mathcal{R}}^2_{\rm GB}=-\frac{48\lambda M^2}{r^{6}}\Big[\frac{3\alpha^2P^4}{M^2 r^6}-\frac{5\alpha P^2}{M r^3} +1\Big].
\end{equation}
\begin{figure*}[t!]
   \centering
  \includegraphics[width=0.4\textwidth]{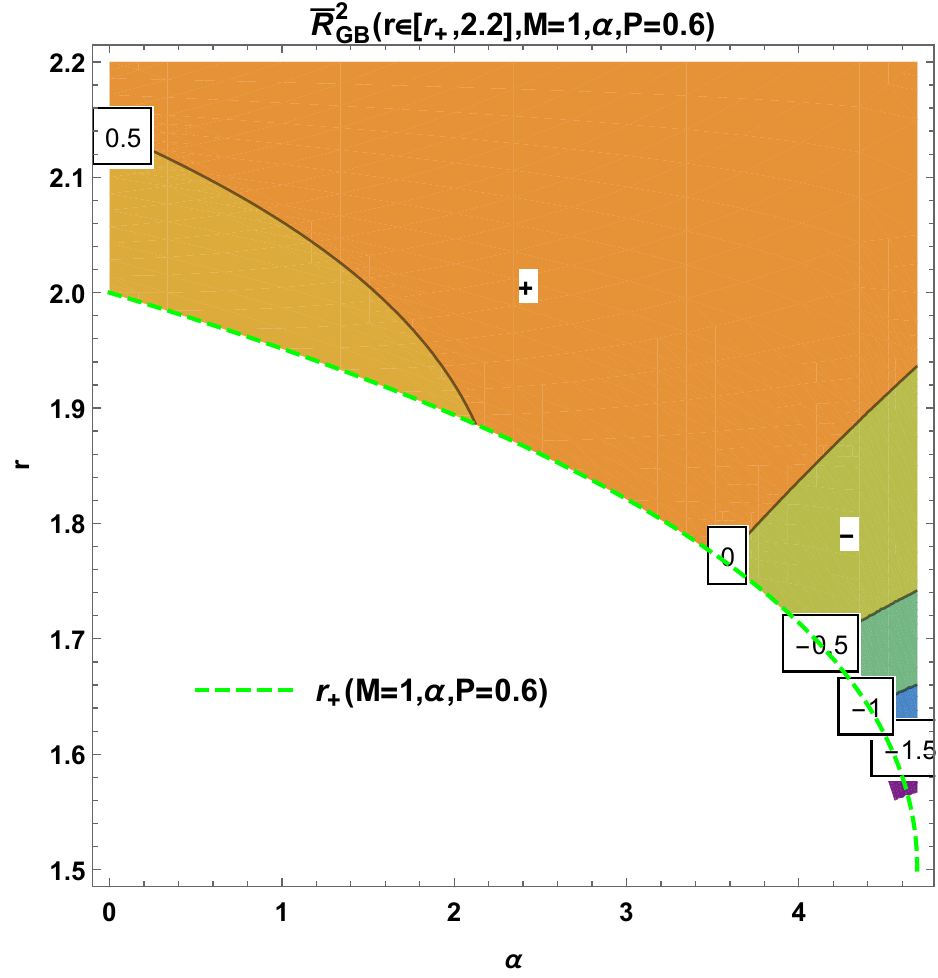}
   \hfill%
\includegraphics[width=0.4\textwidth]{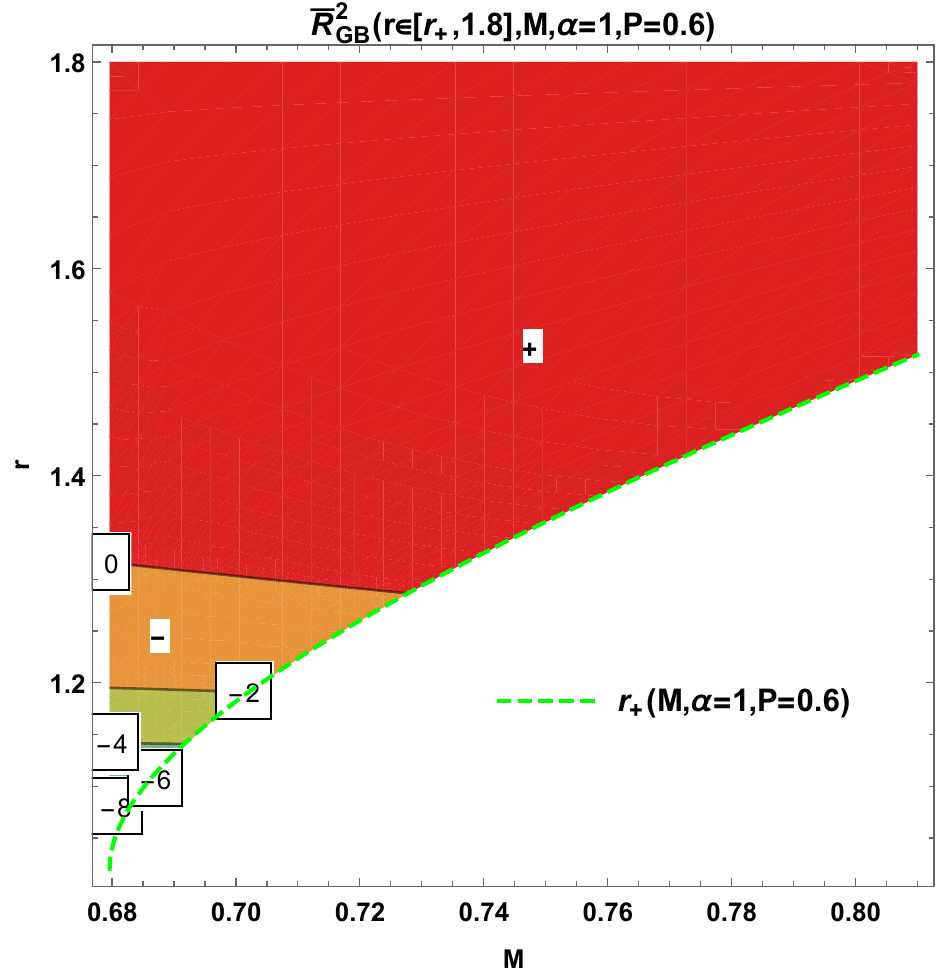}
\caption{ (Left) $\bar{\mathcal{R}}^2_{\rm GB}(r,M=1,\alpha,P=0.6)$  as functions of
$r\in [r_+(M=1,\alpha,P=0.6),2.2]$ and $\alpha\in[0,\alpha_e=4.6875]$ and its zero line is  available, starting from $\alpha=\alpha_c(=3.5653)$ on the horizon.  (Right) $\bar{\mathcal{R}}^2_{\rm GB}(r,M,\alpha=1,P=0.6)$  as functions of
$r\in [r_+(M,\alpha=1,P=0.6),1.8]$ and $M\in[M_{\rm rem}=0.6796,0.81]$ and  its zero line is  available, ending at $M=M_c(=0.7277)$ on the horizon. }\label{fig3}
\end{figure*}
For $\lambda>0$ and $\psi_{00}(t,r_*)\sim u(r_*)e^{-i\omega t}$, GB$^+$ scalarization of Schwarzschild black hole was found for $\alpha=0$~\cite{Doneva:2017bvd,Silva:2017uqg,Antoniou:2017acq}.
For $\lambda<0$, one expects to find  GB$^-$ scalarization with $\alpha\not=0$.
From Fig.~\ref{fig3}, it is clear  that the mass term $m^2_{\rm eff}=-\lambda \bar{\mathcal{R}}^2_{\rm GB}$ could be negative when $\bar{\mathcal{R}}^2_{\rm GB}$ is negative for $\alpha_c\le \alpha \le \alpha_e$ in the near-extremal region and $M_{\rm rem}\le M \le M_c$ in the near-remnant region with $\lambda<0$.

Now, we are in a position to find the  critical onset parameter $\alpha_c$ and mass $M_c$, which determine the lower and upper bounds ($\alpha\ge \alpha_c$, $M\le M_c$) for the  onset of GB$^-$ scalarization  by making use of  the Hod's approach~\cite{Hod:2020jjy}.
To get the critical onset parameters, it is enough to consider the potential term: $V(r)\psi_{00}(t,r_*)=0$.
The onset of  scalarization  is defined by critical black hole which  denotes   the boundary between cqOS-  and  scalarized cqOS-black holes existing  in the limit of $\lambda \to -\infty$.
In this limit, it is characterized by  the presence of a degenerate  binding potential well  whose two turning points
merge at the outer horizon of $r=r_+(M,\alpha,P)$ as
\begin{eqnarray}
 m^2_{\rm eff}(r_+)\psi_{00}(t,r_*)=0. \label{crit-cond}
\end{eqnarray}
From Eq.(\ref{crit-cond}), we find the resonance condition as
\begin{equation}
rc(M,\alpha,P)\equiv \frac{3\alpha^2P^4}{M^2 r^6_+}-\frac{5\alpha P^2}{M r^3_+} +1=0.
\end{equation}
The critical onset action parameter and mass  are  determined by the resonance condition
\begin{eqnarray}
 3\tilde{\alpha}^2-5\tilde{\alpha}+1 =0\label{res-con},\quad \tilde{\alpha}=\frac{\alpha P^2}{M r_+^3}.
\end{eqnarray}

\begin{figure*}[t!]
   \centering
  \includegraphics[width=0.6\textwidth]{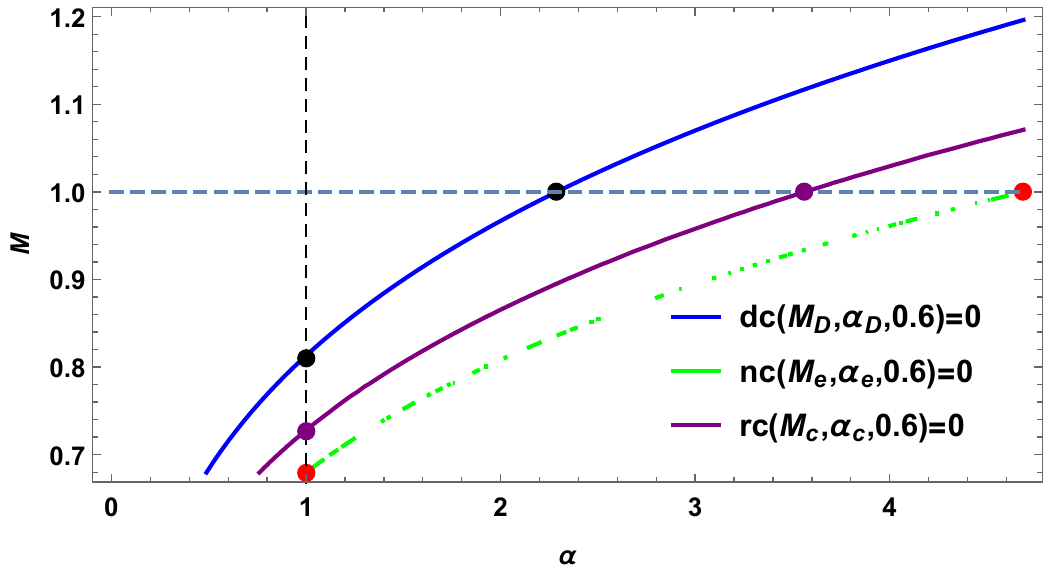}
\caption{ Three different curves for $nc(M_e,\alpha_e,0.6)=0$, $dc(M_D,\alpha_D,0.6)=0$, and $rc(M_c,\alpha_c,0.6)=0$ for $M\in[0.6796,1.2]$ and $\alpha\in[0,4.6875]$, including  extremal/remnant points (red dot), two Davies points (black dot), and two resonance  points (purple dot) for $M=1$ and $\alpha=1$. }\label{fig4}
\end{figure*}

Solving $\tilde{\alpha}=0.2324$ for $\alpha=\alpha_c$ with $M_c=1$ and $P=0.6$, and for $M=M_c$ with $\alpha_c=1$ and $P=0.6$ leads to
the critical onset  mass and action parameter  for  GB$^-$ scalarization as
\begin{eqnarray}
(M_c=1,\alpha_c=3.5653)\quad {\rm and} \quad (M_c=0.7277,\alpha_c=1).
\end{eqnarray}
 Fig.~\ref{fig4} shows  that $\{dc(M_D,\alpha_D,0.6)=0\} \not=\{rc(M_c,\alpha_c,0.6)=0\}$, implying that Davies curve cannot be consistent with  critical onset (resonance) curve.
Observing Fig.~\ref{fig2}, $\alpha = \alpha_c$ is regarded as the lower bound for GB$^-$ scalarization ($\alpha_c\le \alpha \le \alpha_e,~C>0$), while  $M= M_c$ is considered as the upper bound for GB$^-$ scalarization ($M_{rem}\le M \le M_c,~C>0$).  It is worth noting that the presence of such restrictions may lead to  the single branch of scalarized cqOS-black holes.

\begin{figure*}[t!]
   \centering
    \mbox{
   (a)
  \includegraphics[width=0.43\textwidth]{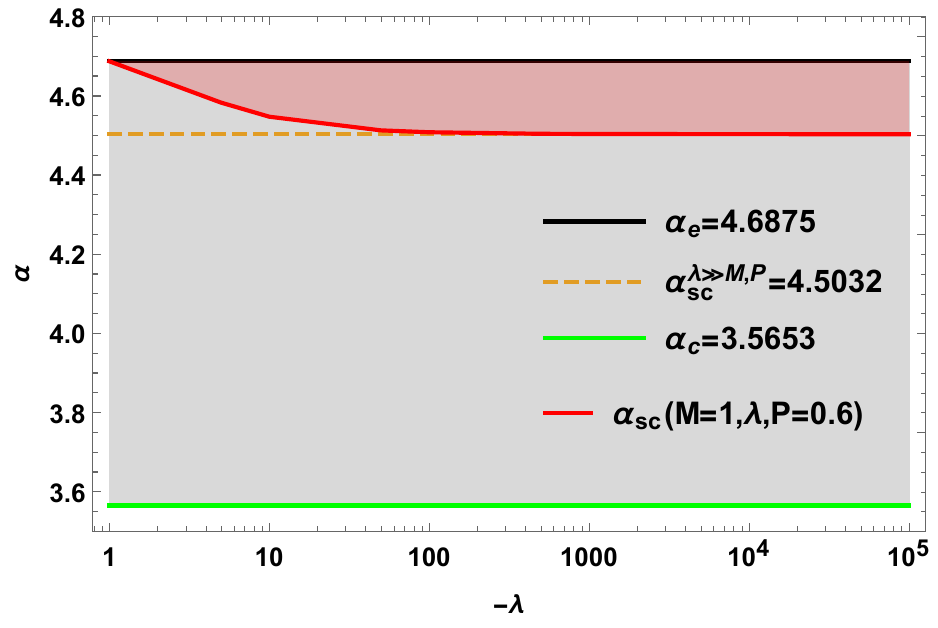}
 (b)
    \includegraphics[width=0.43\textwidth]{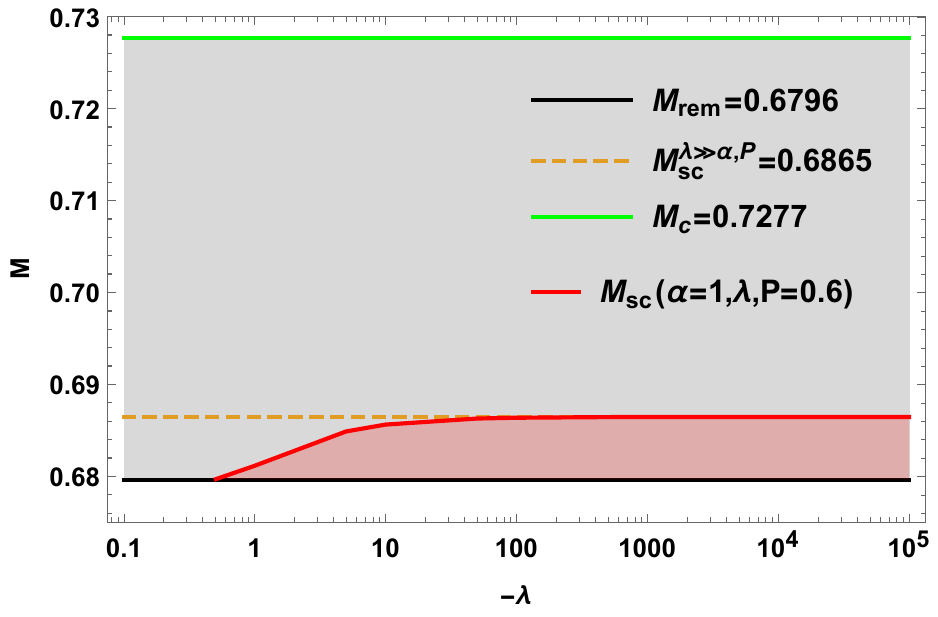}}
\caption{(a) Sufficient condition of $\alpha_{\rm sc}(M=1,\lambda,P=0.6)$ with the critical onset parameter $\alpha=\alpha_c$ as the lower bound.  A dashed line denotes the sufficient condition $\alpha=\alpha_{\rm sc}^{-\lambda\gg M,P}$. A top line denotes the extremal point at $\alpha=\alpha_e$ as the upper bound.
(b) Graph for $M_{\rm sc}(\alpha=1,\lambda,P=0.6)$ with  the critical onset mass $M=M_c$ as the upper bound  and the sufficient condition $M=M_{\rm sc}^{-\lambda\gg  \alpha,P}$.  A bottom line represents the remnant point at $M=M_{\rm rem}$ as the lower bound.}\label{fig5}
\end{figure*}

On the other hand, the sufficient condition for the instability can be obtained by imposing $\int^\infty_{r_+}V(r)dr/g(r) <0$~\cite{Dotti:2004sh}.
For $-\lambda \gg M,\alpha,P$,  this condition can reduce to
\begin{equation}
\int^\infty_{r_+}m^2_{\rm eff}(r) dr \equiv I<0,
\end{equation}
which leads to
\begin{equation}
I=\frac{6(120 \alpha^2 P^4-275  \alpha  P^2 M r_+^3+88 M^2r_+^6)}{55r_+^{11}}<0.
\end{equation}
We solve  $I=0$ to find two sufficient conditions: $\alpha_{sc}^{-\lambda\gg M,P}=4.5032$ for $M=1,~M_{sc}^{-\lambda\gg  \alpha,P}=0.6865$ for $\alpha=1$.

To find  $\alpha_{\rm sc}(M=1,\lambda,P=0.6)$ and $M_{\rm sc}(\alpha=1,\lambda,P=0.6)$, we analyze the condition of  $\int^\infty_{r_+}V(r)dr/g(r) <0$  numerically
for a given $\lambda$.
As is depicted in Fig.~\ref{fig5}, we find that  $\alpha_{\rm sc}(1,\lambda,0.6)$ is a decreasing function, connecting between  $\alpha_e$ and  $\alpha_{\rm sc}^{-\lambda\gg M,P}$, while $M_{\rm sc}(1,\lambda,0.6)$ is an increasing function, connecting between $M_{\rm rem}$ and
$M_{\rm sc}^{-\lambda\gg \alpha,P}$.  We note that two red-shaded regions represent sufficiently unstable region for GB$^-$ scalarization.

Finally, to obtain  the threshold of instability [$\alpha_{\rm th}(M=1,\lambda,P=0.6)$ and $M_{\rm th}(\alpha=1,\lambda,P=0.6)$] which can connect from ($\alpha_c,M_{\rm min}$) to ($\alpha_e,M_{c}$), one has to solve Eq.(\ref{mode-d}) to find its time-dependent behaviors   by utilizing the fourth-order Runge-Kutta method for time evolution and
considering an initial Gaussian wave-packet of $\psi_{0,0}(0,r_*)=e^{-(r_*-r_*^c)^2/\lambda}$~\cite{Lai:2022spn,Lai:2022ppn,Chen:2025wze}. For a given $\lambda$, its time-independent behavior  determines the threshold of instability [$\alpha_{\rm th}(M=1,\lambda,P=0.6)$ and $M_{\rm th}(\alpha=1,\lambda,P=0.6)$] for GB$^-$ scalarization.
However,  we realize that  this task  will  amount to the other project.

\section{Scalarized charged black holes}
Instead of finding the thresholds of instability $\alpha_{\rm th}(1,\lambda,0.6)$,
we directly obtain scalarized cqOS-black holes by solving the full equations.
For this purpose, we adopt the following metric and field ansatz~\cite{Doneva:2017bvd,Doneva:2018rou}:
\begin{eqnarray}
ds^2= -A(r)dt^2 + \frac{dr^2}{B(r)} + r^2 (d\theta^2 + \sin^2\theta d\phi^2 )~,\quad \phi=\psi(r),\quad A=A_\varphi.
\end{eqnarray}
Substituting the gauge field ansatz into the Maxwell equation yields the magnetic  potential $A_{\varphi}=-P \cos \theta$, which in turn determines  $F_{\theta \varphi}=P\sin\theta$ and $\mathcal{F}=2P^2/r^4$.
This implies that an approximate solution for $A_{\varphi}$ is unnecessary and thus,  the nonlinear-Maxwell equation is solved exactly in this case.
Then, two  equations (\ref{equa1}) and (\ref{s-equa}) reduce to the forms

\begin{eqnarray}
&&-\frac{6\alpha P^2}{r^4}+\frac{rB'}{2}+\Big(\frac{r^2B}{2}-8\lambda r B^2\psi \psi'\Big)\frac{A''}{A}+\Big(-\frac{r^2 B}{4}+4\lambda r B^2 \psi \psi'\Big)\frac{A'^2}{A^2} \nonumber \\
&&+\Big(\frac{rB}{2}+B'(\frac{r^2}{4}-12 \lambda r B \psi \psi')-8\lambda r B^2( \psi'^2+\psi \psi'')\Big)\frac{A'}{A}+r^2B \psi'^2=0,\label{DRFE1} \\
&&1-\frac{3\alpha P^2}{r^4}-B'(r+8\lambda \psi\psi')-B\Big(1+(r^2+16\lambda)\psi'^2-8\lambda \psi(3B'\psi'-2\psi'')\Big) \nonumber \\
&& \hspace{3cm}+16\lambda B^2(\psi'^2+\psi \psi'')=0, \label{DRFE2} \\
&& \psi'' + \frac{1}{2}\left(\frac{4}{r}+\frac{A'}{A} +\frac{B'}{B}\right)\psi'+\frac{2\lambda \psi}{r^2A^2B}\Big(A'[-(-1+B)BA'  \nonumber \\
&&\hspace{3cm} +A(-1+3B)B']+2A(-1+B)BA''\Big) =0. \label{DRFE3}
\end{eqnarray}
Considering the existence of outer horizon located at $r=r_+$, the near-horizon solution to Eqs.~\eqref{DRFE1}-\eqref{DRFE3} can be expressed as~\cite{Antoniou:2017acq}
\begin{eqnarray}
&&A(r)=A_1(r-r_+)+A_2(r-r_+)^2\cdots,  \label{aps-0} \\
&& B(r)=B_1(r-r_+)+B_2(r-r_+)^2+\cdots, \label{aps-1} \\
&&\psi(r)=\psi_0+\psi_1(r-r_+)+\cdots.  \label{aps-2}
\end{eqnarray}
Here, the constant scalar $\psi_0= \psi(r_+)$ at the horizon  can  be
determined when matching with an asymptotically flat solution in the far-region
\begin{eqnarray}\label{bdyinf}
\{A,B\}=1+\sum_n^\infty \frac{\{p_n,q_n\}}{r^n},\quad \psi=\psi_\infty+\sum_n^\infty \frac{d_n}{r^n}
\end{eqnarray}
with coefficients of $p_n,~q_n,$ and $d_n$ to be determined.  A choice of $\psi_\infty$ depends on the sign of coupling constant $\lambda$.

\subsection{GB$^{+}$ scalarization with $\alpha=0$ and $\lambda>0$}

For $\alpha=0$ and $\lambda>0$,  this model  reduces to carrying out the GB$^{+}$ scalarization for Schwarzschild black hole in the EGBS theory.  With the positive coupling parameter $\lambda$, one may find the usual GB$^+$ scalarization of Schwarzschild black hole, giving infinite branches of scalarized black holes from infinite scalar clouds~\cite{Myung:2018iyq}.  Here, its branches are determined by $\frac{r_+}{2\sqrt{\lambda_n}}=1.174(n=0),~0.453(n=1),~0.280(n=2),~0.202(n=3)$. For  the fundamental ($n=0$) branch, its starting point is given by  $\lambda_0=1.0188$ for $r_+=2.37$. Hence, its scalarized black hole solutions are allowed for $\lambda\ge \lambda_0$ in the fundamental branch. 

\begin{figure*}[t!]
   \centering
    \mbox{
   (a)
  \includegraphics[width=0.45\textwidth]{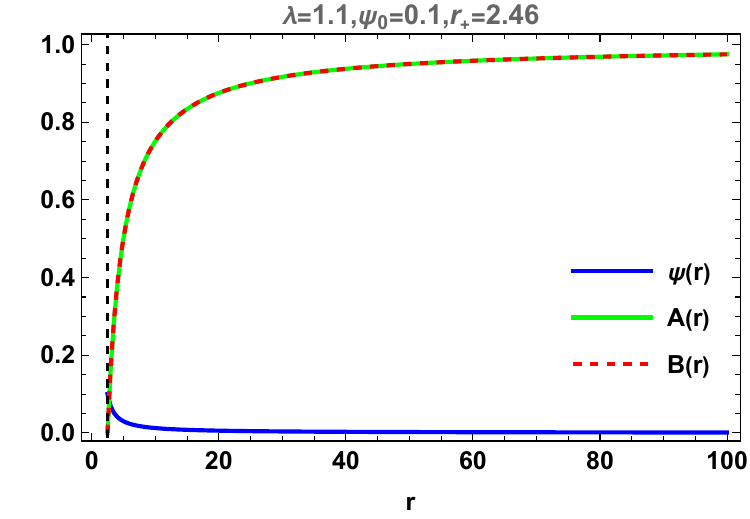}
 (b)
    \includegraphics[width=0.45\textwidth]{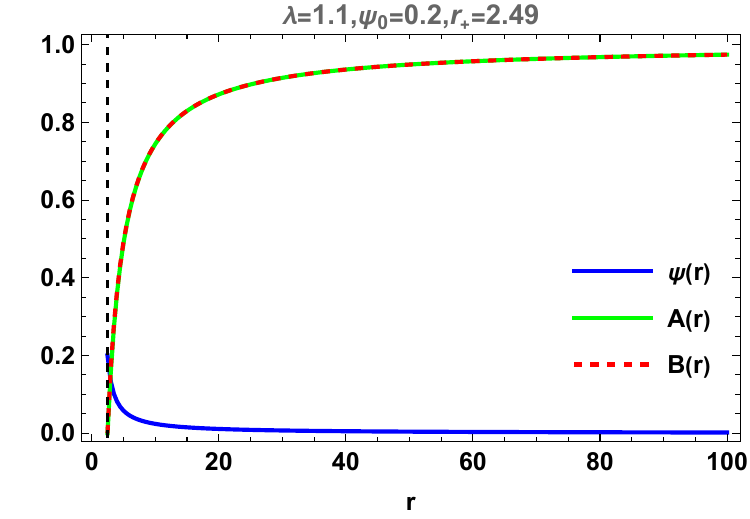}}
\caption{Scalarized  black holes obtained from  GB$^+$ scalarization with $P=0,\alpha=0,\lambda=1.1$ for (a) $\psi_0=0.1$ and (b) $\psi_0=0.2$ in the fundamental branch.}\label{fig6}
\end{figure*}

To obtain  the scalarized black hole solution in the $n=0$ branch, we wish to set $\psi_\infty=0$ and $A(\infty)=B(\infty)$ as the shooting condition to solve Eqs.~\eqref{DRFE1}-\eqref{DRFE3} numerically.  For $\lambda=1.1>\lambda_0$, we select $\psi_0=0.1$ and 0.2 to represent the radial profiles for metric functions and scalar field in Fig.~\ref{fig6}.  It is observed  that the scalar field decays monotonically and rapidly approaches zero as $r$ increases. Meanwhile, the metric functions $A(r)$ and $B(r)$ are zero at the horizon, then increase monotonically, approach 1 at infinity, and nearly coincide with each other throughout radial space. 
Although the radial profiles of the metric functions for different $\psi_0$ exhibit very similar qualitative behavior, we note that these solutions correspond to different black hole configurations. 
In particular, our calculations show that, that for $\psi_0=0.1$, the black hole mass is $M=1.24$ and the Wald entropy is $S_W=19.4$, while at $\psi_0=0.2$, the black hole mass is $M=1.28$ and the Wald entropy is $S_W=20.6$.
They vary mildly with $\psi_0$, as indicated above the figure.
The differences remain quantitatively small for the parameter range considered here, which explains why the profiles appear nearly indistinguishable.

Finally, we mention the case of  $\lambda>0$ and $\alpha>0$.
Its scalarization corresponds to GB$^+$ scalarization of charged qOS-black holes which is simliar to GB$^+$ scalarization of Reissner-Nordstr\"om black holes appeared in~\cite{Brihaye:2019kvj}. Very similar to the previous uncharged case, charged qOS-black holes undergo spontaneous scalarization for sufficiently large scalar-tensor coupling $\lambda$–a phenomenon attributed to a tachyonic instability of the scalar field system.

\subsection{GB$^{-}$ scalarization with $\alpha\not=0$ and $\lambda<0$}

To find scalarized cqOS-black holes in  the single branch, we choose $P=0.6$ and $\alpha=4.6$ which belongs to $\alpha_{\rm sc}(M=1,\lambda=-10,P=0.6)(=4.5473)<\alpha<\alpha_e(M=1,P=0.6)(=4.6875)$.
In the case of negative coupling $\lambda<0$, we find a non-monotonically numerical solution for the scalar field. 
It is worth noting that the scalar field exhibits a monotonically increasing behavior for relatively large horizon radii. 
However, by tuning the model parameters, one can also obtain physically admissible solutions in which the scalar field decays monotonically, as well as excited-state solutions with node number $n > 0$. 
Accordingly, we impose the shooting conditions $\psi_{\infty} = 0$ together with $A(\infty)=B(\infty)$ in order to construct viable scalarized black hole solutions.
Here, $\psi_0$ is regarded as a free parameter that can be adjusted. 
However, across various coupling scenarios, all numerical solutions we can find indicate that the scalar constant $\psi_0$ cannot grow very large, implying that the constant scalar $\psi_0$ possessed by scalarized cqOS black holes is quite small.
At the level of the first-order approximation, the black hole mass can be extracted from the coefficient $p_1$, while the scalar charge is determined by $d_1$. 
For convenience, we redefine the parameters as $M = p_1, q_s = d_1$.

\begin{figure*}[t!]
   \centering
    \mbox{
   (a)
   \includegraphics[width=0.28\textwidth]{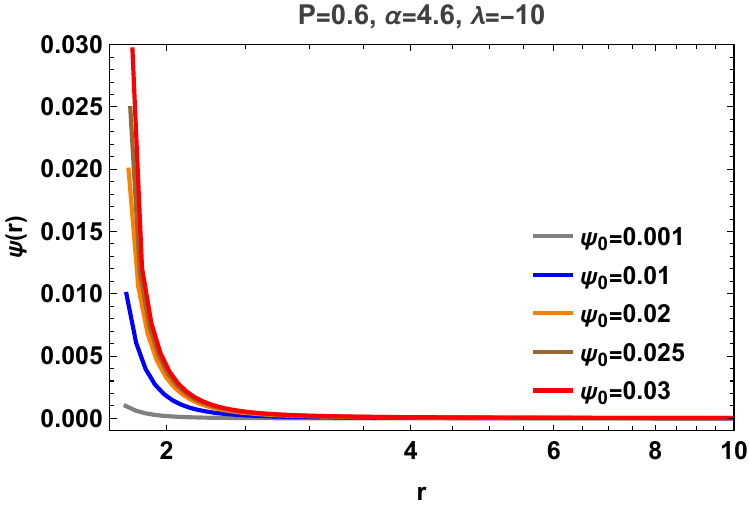}
   (b)
   \includegraphics[width=0.28\textwidth]{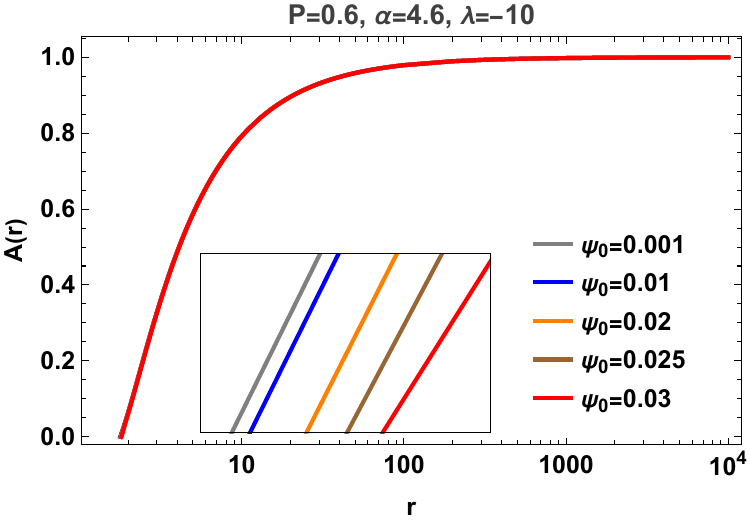}
   (c)
   \includegraphics[width=0.28\textwidth]{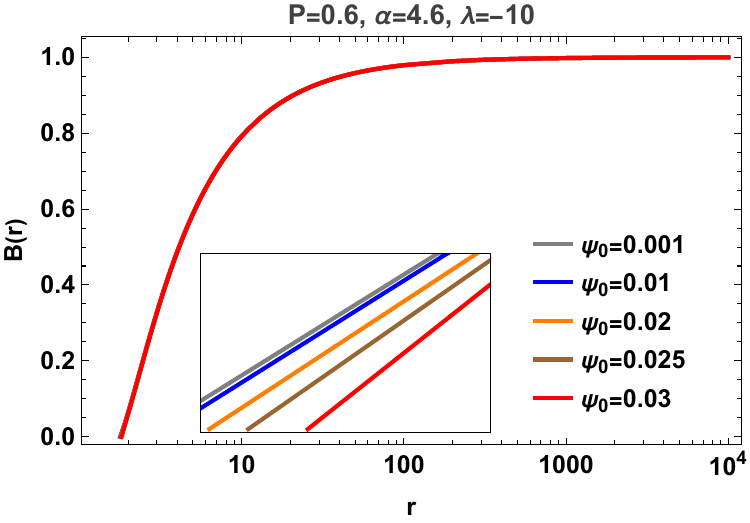}}
\caption{Scalarized cqOS black holes obtained from GB$^-$ scalarization with $P=0.6,\alpha=4.6,\lambda=-10$. (a) Scalar  $\psi(r)$ with $\psi_\infty=0.1$. (b) Metric function  $A(r)$. (c) Metric function $B(r)$.}\label{fig7}
\end{figure*}

For a choice of $\lambda=-10$, the scalar profile $\psi(r)$ and two metric functions $A(r)$ and $B(r)$ with different $\psi_0$ are shown in Fig.~\ref{fig7}. 
An interesting observation is that, although both the GB$^+$ and GB$^-$ scalarization branches admit solutions in which the scalar field decays monotonically, the decay in the GB$^-$ case is significantly more rapid. 
As a result, the scalar field becomes effectively negligible throughout the vast majority of the exterior region of the black hole spacetime.
In conjunction with the fact that the scalar hair is generally difficult to amplify to large amplitudes, this implies that, in the GB$^-$ scenario, it is challenging to sustain sufficiently pronounced scalar hair.


\begin{figure*}[t!]
   \centering
   \mbox{
   (a)
   \includegraphics[width=0.45\textwidth]{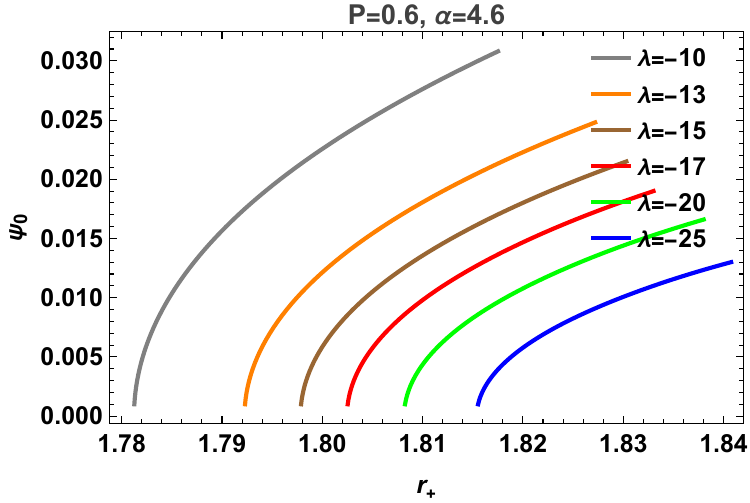}
   (b)
   \includegraphics[width=0.45\textwidth]{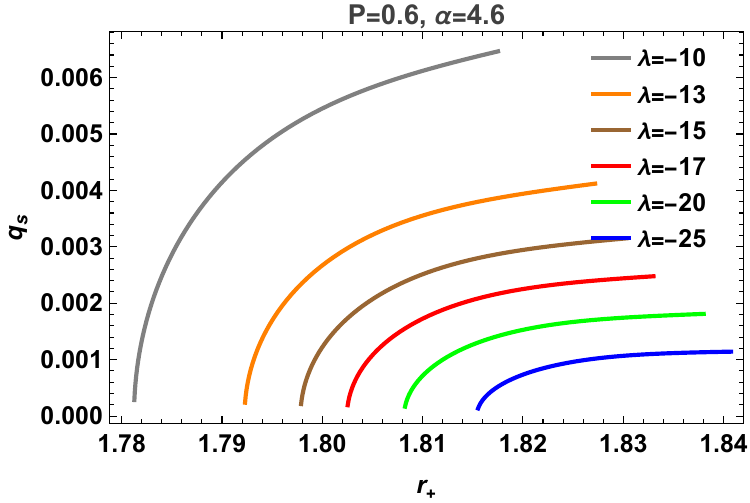}}
\caption{(a) Scalar hair $\psi_0$ and (b) scalar charge $q_s$ as the function of horizon radius $r_+$. }\label{fig8}
\end{figure*}

Two metric functions $A(r)$ and $B(r)$ satisfy the conditions for the existence of a black hole. It approaches zero at the event horizon, then increases monotonically until it tends toward 1 at infinity.  
A notable feature is observed that the horizon radius increases approximately monotonically as the scalar hair becomes stronger. This trend is also consistent with the behavior of the scalar field shown in panel (a).
To clarify this behavior more explicitly, we plot the scalar hair $\psi_0$ and the scalar charge $q_s$ as functions of the horizon radius $r_+$ in Fig.~\ref{fig8}.
It can be seen that both the scalar hair and the scalar charge grow monotonically with increasing horizon radius. 
However, for the scalar charge, one can clearly observe a tendency to approach a constant value, a behavior that is commonly encountered in previous studies of spontaneous scalarization. 
Moreover, for larger coupling strength, the scalarized black holes exhibit larger horizon radii, while the allowed ranges of the scalar hair and scalar charge become suppressed. This indicates that the scalar hair becomes progressively less pronounced as the coupling strength increases.


\begin{figure*}[t!]
   \centering
    \mbox{
   (a)
    \includegraphics[width=0.28\textwidth]{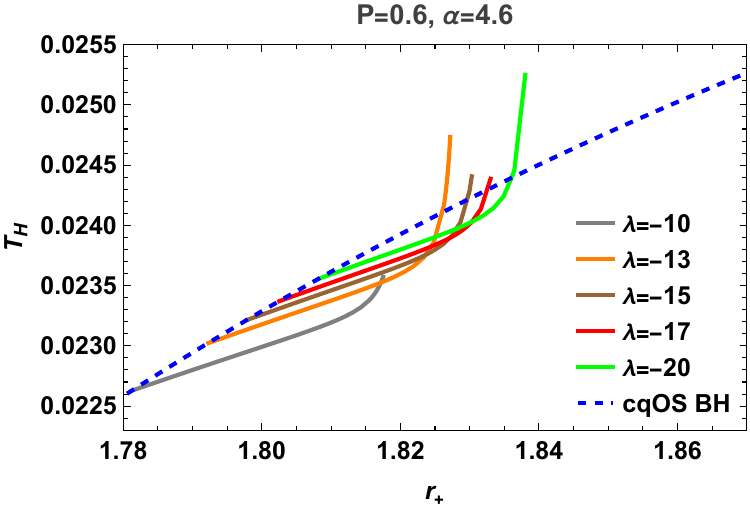}
   (b)
    \includegraphics[width=0.28\textwidth]{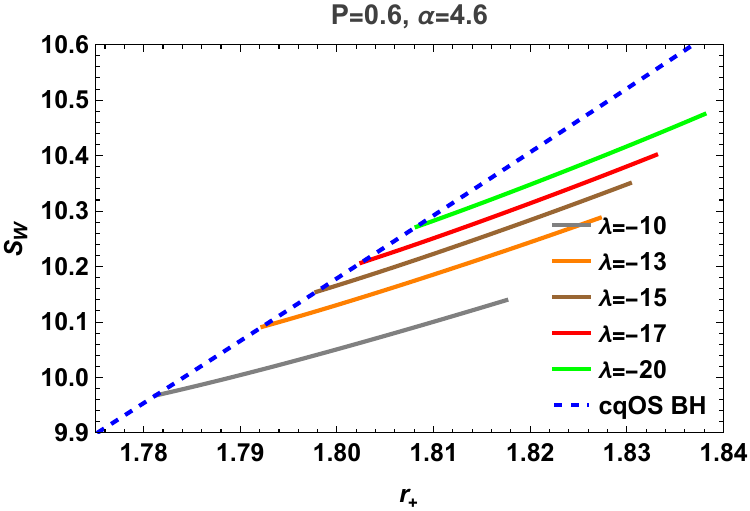}
   (c)
    \includegraphics[width=0.28\textwidth]{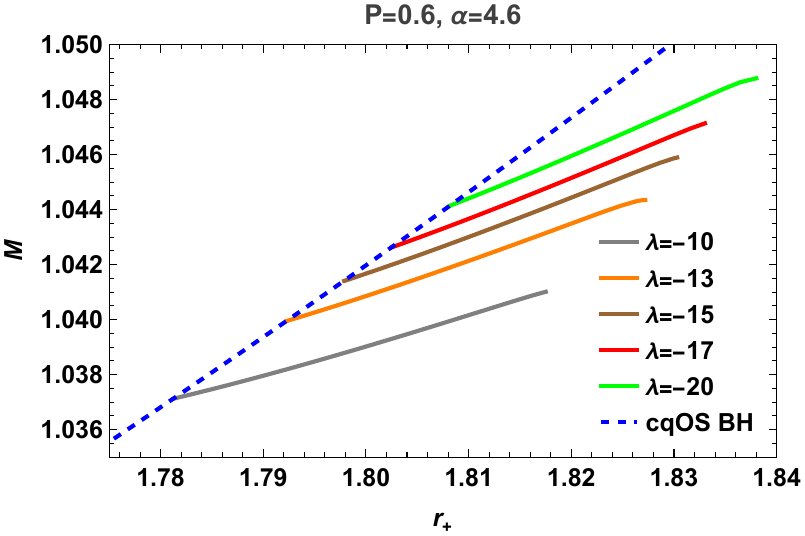}}
\caption{(a) Hawking temperature $T_H$, (b) Wald entropy $S_W$ and (c) black hole mass $M$  as  functions  of the event horizon radius $r_+$ for different coupling constants ($\lambda=-10\sim -20$).}\label{fig9}
\end{figure*}

The emergence of scalarized black holes directly influences the behavior of the thermodynamic quantities. So we further observe how the temperature and entropy behave according to the horizon radius $r_+$ and coupling constant $\lambda$. 
The Hawking temperature is defined as 
\begin{equation}
    T_H (r_+)= \frac{1}{4\pi} \sqrt{A'(r_+) B'(r_+)}.
\end{equation}
Due to the  GB term, the black hole entropy no longer follows the  area-law. Instead, we  compute the Wald entropy as 
\begin{equation}
    S_{\text{W}} = \frac{a_H(r_+)}{4} + 4\pi f(\psi_0),
    \label{eq:wald_entropy}
\end{equation}
where $a_H(r_+) = 4\pi r_+^2$ denotes the horizon area.

We wish to investigate the numerical solutions under different coupling constants. Fig.~\ref{fig9} illustrates the Hawking temperature, Wald entropy and black hole mass as functions of horizon radius $r_+$.
The blue dashed curves correspond to the hairless cqOS black hole solutions. One can see that all scalarized black hole branches bifurcate from the corresponding hairless cqOS configurations.
As for the temperature, it is evident that, in general, scalarized black holes have lower temperatures than their cqOS counterparts. 
However, near the endpoint of the scalarization branch, the temperature exhibits a sharp increase. 
This region coincides with the regime shown in Fig.~\ref{fig8}(b), where the scalar charge approaches a constant value.
In contrast to the temperature, both the Wald entropy and the black hole mass display an approximately linear dependence along the scalarized branches, and are systematically lower than those of the corresponding hairless cqOS solutions.
This indicates that the emergence of scalar hair has a more pronounced effect on the black hole temperature than on other thermodynamic quantities.
Moreover, the numerical results, seen in Fig.~\ref{fig9}(b) and (c), indicate that there is a monotonic relationship between Wald entropy and black hole mass.
In the limit of vanishing scalar hair, the solutions smoothly reduce to the cqOS black hole.

\section{Stability analysis for scalarized charged black hole}

Before we proceed, it was shown for GB$^+$ scalarization that for an exponential coupling of $(1-e^{-6 \phi^2})/12$,
its fundamental branch of scalarized black holes is stable against the radial perturbation~\cite{Doneva:2017bvd,Silva:2017uqg}, while  the fundamental branch of scalarized black holes is unstable against the radial perturbation for a quadratic coupling of $ \phi^2$~\cite{Blazquez-Salcedo:2018jnn}.  The latter branch  is unstable because its width is short and it terminates at some nonzero mass due to  violation of the regularity condition for $r_+$.
Hence, we are  very curious to study the stability of scalarized black holes in the whole single branch obtained from GB$^-$ scalarization with the quadratic coupling $f(\phi)=2\lambda \phi^2$.

For this purpose, let us consider a radial perturbation around the scalarized black holes as 
\begin{align}
    & ds^2= -\big(A(r)+\epsilon H_0(t,r)\big)dt^2 + \frac{dr^2}{B(r)+\epsilon H_1(t,r)} + r^2 (d\theta^2 + \sin^2\theta d\phi^2 )~,\nonumber\\
    & \phi=\psi(r)+\epsilon \Phi(t,r).
\end{align}
Here, $H_0(t,r)$, $H_1(t,r)$ and $\Phi(t,r)$ denote the perturbations propagating on the numerical background, while $\epsilon$ represents a small deviation from our scalarized black holes.
It is important to note that no perturbations are introduced for the gauge field $A_{\varphi}$.
Henceforth, we focus on the $l=0$($s$-mode) scalar propagation, with higher angular momentum modes ($l\ne0$) being disregarded. 
In this case, the two perturbed metric fields $H_0(t,r),H_1(t,r)$ become redundant via a decoupling procedure, leaving a single linearized scalar equation~\cite{Minamitsuji:2018xde}
\begin{align}
    -\rho_0 \ddot{\Phi}(t,r) + \rho_0 \Phi''(t,r) + \rho_1 \Phi'(t,r) + \rho_2 \Phi(t,r)=0,
\end{align}
where three coefficients ($\rho_0,\rho_1,\rho_2$) are given in Appendix~\ref{appendix-A}.
By assuming $\Phi=C(r)R(r)$, the scalar perturbation reduces to a Schr\"odinger-type equation
\begin{align}\label{pertub}
    \frac{d^2R(t,r_*)}{dr_*^2}-V_{\text{eff}}(r)R(t,r_*)=A(r)B(r)\frac{d^2R(t,r_*)}{dt^2},
\end{align}
with the tortoise coordinate $r_*$ defined by $dr_*=\frac{dr}{\sqrt{A(r)B(r)}}$.
The auxiliary function $C(r)$ satisfies the condition
\begin{align}
    \frac{C'(r)}{C(r)}=\frac{1}{4}[\ln(A(r)B(r))]'-\frac{\rho_1}{2\rho_0}.
\end{align}
The effective potential takes the form
\begin{align}\label{potential}
    V_{\text{eff}}(r)=-A(r)B(r)\bigg[\frac{1}{4}(\ln{A(r)B(r)})''+\frac{1}{16}((\ln{A(r)B(r)})')^2-\frac{1}{2}(\frac{\rho_1}{\rho_0})'-\frac{\rho_1^2}{4\rho_0^2}+\frac{\rho_2}{\rho_0}\bigg].
\end{align}

\begin{figure*}[t!]
   \centering
    \mbox{
(a)
  \includegraphics[width=0.45\textwidth]{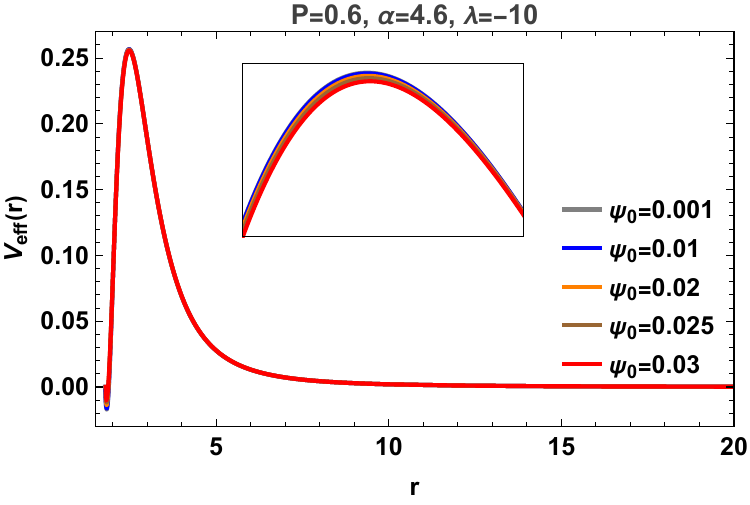}
(b)
    \includegraphics[width=0.45\textwidth]{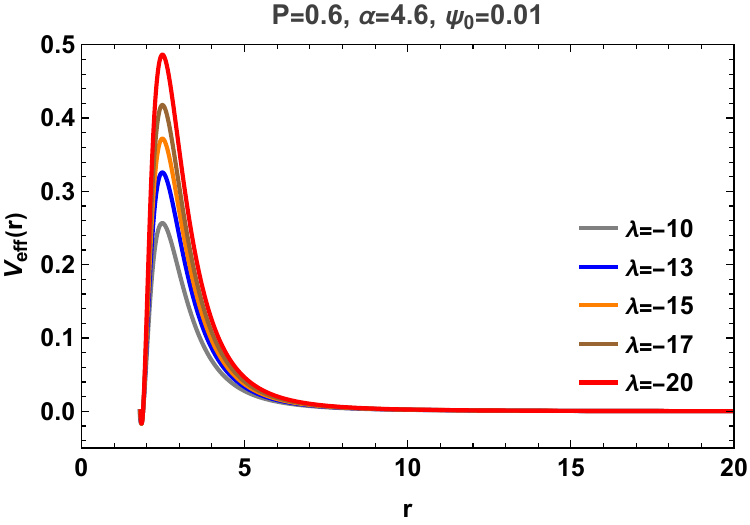}}
\caption{Radial profiles of effective potential under scalar perturbation for different (a) scalar constant $\psi_0$ and (b) coupling constants $\lambda$ }\label{fig10}
\end{figure*}

Fig.~\ref{fig10} displays the radial profiles of effective potential for the $s$-mode ($\ell = 0$) scalar perturbations under different values of $\psi_0$ and the coupling constant $\lambda$. 
As shown in this figure,  the effective potential develops a pronounced potential barrier outside the event horizon.
Moreover, one observes that the height of the effective potential barrier decreases slightly as the scalar hair increases. 
In contrast, an increase in the coupling strength significantly enhances the height of the potential barrier.

\begin{figure*}[t!]
   \centering
    \mbox{
(a)
  \includegraphics[width=0.45\textwidth]{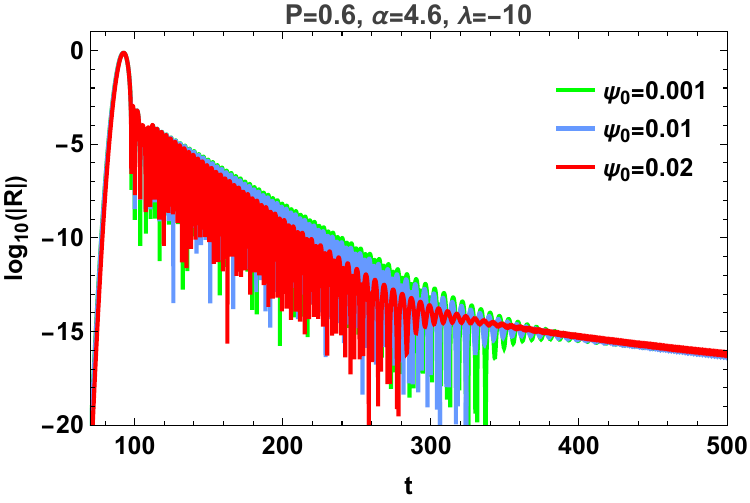}
(b)
    \includegraphics[width=0.45\textwidth]{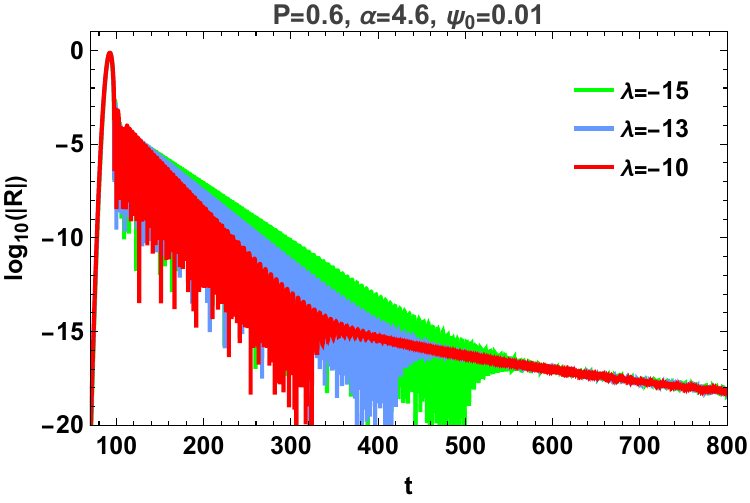}}
\caption{Time evolution of the scalar perturbation for different (a) scalar hair $\psi_0$ and (b) coupling constants $\lambda$. }\label{fig11}
\end{figure*}

It should be noted that the effective potential develops a shallow potential well near the event horizon. 
Nevertheless, the sufficiently high potential barrier strongly suggests the stability of the scalarized cqOS-black holes solution under scalar perturbations. 
To further substantiate this conclusion, one possible approach is to compute the quasinormal mode (QNM) spectrum. 
However, the non-standard form of the perturbation equation~\eqref{pertub}, together with the highly complicated form of the effective potential, makes the QNM calculation extremely challenging. 
As an alternative, one may instead investigate the time-domain evolution of scalar perturbations.

To this end, we discretize the perturbation equation~\eqref{pertub} as
\begin{align}\label{discre}
    \frac{R_{i,j+1}-2R_{i,j}+R_{i,j-1}}{\Delta r_*^2}-V_jR_{i,j}=A_jB_j\frac{R_{i+1,j}-2R_{i,j}+R_{i-1,j}}{\Delta t^2}+\mathcal{O}(\Delta t^2)+\mathcal{O}(\Delta r_*^2),
\end{align}
by defining $R(t,r_*)=R(i\Delta t,j\Delta r_*)=R_{i,j}$, $A(r(r_*))=A(j\Delta r_*)=A_j$, $B(r(r_*))=B(j\Delta r_*)=B_j$ and $V_{\text{eff}}(r(r_*))=V_{\text{eff}}(j\Delta r_*)=V_j$.
In this case, we consider the initial Gaussian wave packet $R(t=0,r_*)=\exp(-\frac{(r_*-a)^2}{2b^2})$ with $a=10$ and $b=4$, $\Delta t/\Delta r_* =0.45$ is set to satisfy the Neumann stability condition $\Delta t/\Delta r_*<1$.
Another important point is that both A and B vanish at the event horizon. As a consequence, the time-domain evolution of Eq.~\eqref{discre} becomes divergent near the horizon during the numerical integration. 
To avoid this issue, we start the numerical calculation slightly outside the horizon, namely at $r_{\rm start}=r_{+}+0.002$.

The corresponding numerical results are presented in Fig.~\ref{fig11}.
It can be observed that, the scalar perturbation undergoes an exponential decay at early times and subsequently evolves into a late-time power-law tail.
Moreover, the left panel shows that increasing the scalar hair leads to a faster decay of the perturbation and an earlier onset of the power-law tail. 
In contrast, the right panel indicates that a larger value of the coupling parameter causes the perturbation to decay more slowly and delays the transition to the late-time tail behavior.
Consequently, since the perturbation remains decaying throughout the entire time evolution, the results confirm the stability of the scalarized black hole solutions from GB$^-$ scalarization under scalar perturbations.

\section{Discussions}

We have investigated GB$^-$ scalarization of cqOS-black holes  obtained from  the EGBS-NED theory.
An important feature of these black holes is that they are  described  by mass $M$ and  action parameter $\alpha$ with a fixed  magnetic charge $P$.
This means that the magnetic charge $P$  plays a subsidiary role in studying cqOS-black holes, contrasting with action parameter $\alpha$.

We have studied  thermodynamics of cqOS-black holes to obtain  peculiar points of extremal point ($\alpha_e$), Davies points ($\alpha_D,~M_D$), remnant point ($M_{\rm rem}$), and allowed regions for the  mass ($M\in[M_{\rm rem},\infty]$) and the action  parameter ($\alpha\in[0,\alpha_e]$).
Thermodynamic study was done   by finding thermodynamic quantities, checking  the first-law thermodynamics and the Smarr formula, and recognizing   a  phase transition at Davies points  in the heat capacity.

Introducing $f(\phi)=2\lambda \phi^2$ scalar coupling to GB term, we found onset of GB$^-$  scalarization with critical onset parameters ($\alpha_c,~M_c$), implying that the single branch of scalarized cqOS-black holes exists.
It is worth noting  that the allowed unstable regions are narrow as  $\alpha_{th}(M=1,P=0.6, \lambda)\in[\alpha_c,\alpha_e]$ and $M_{th}(\alpha=1,P=0.6,\lambda)\in[M_{\rm rem},M_c]$ for GB$^-$ scalarization.
To this model, we did not find any close connection between thermodynamics  and  onset of GB$^-$ scalarization
because the Davies curve  obtained  from  heat capacity has noting to do  with  the critical onset curve. This reflects that  the nature of the present model is considered  as charged quantum Oppenheimer-Snyder model~\cite{Mazharimousavi:2025lld}, but not quantum Oppenheimer-Snyder model~\cite{Lewandowski:2022zce}.

We have obtained all scalarized cqOS-black holes through GB$^\pm$ scalarizations  by solving Eqs.(\ref{equa1}) and (\ref{s-equa}) numerically.
In the case $\alpha=0$ and $\lambda>0$, which corresponds to GB$^+$ scalarization, our model reduces to the standard EGBS scenario for the Schwarzschild black hole.
In contrast, for GB$^-$ case with $\alpha\neq 0$ and $\lambda<0$, a single branch of scalarized solutions is found in a narrow parameter space.
It is worth noting that, in the GB$^-$ scalarization branch, the scalar field decays much more rapidly compared to the GB$^+$ case.
From a thermodynamic perspective, the temperature, Wald entropy, and mass of the scalarized black holes all bifurcate from their counterparts in the hairless cqOS solutions, and are generally slightly smaller than those of the corresponding hairless black holes.

Finally, we have examined linear stability of scalarized cqOS-black holes under scalar perturbations.  
We found that the effective potential possesses a sufficiently high barrier structure, while the time-domain evolution of the scalar perturbation remains continuously damped, exhibiting the characteristic behavior of exponential decay followed by a late-time power-law tail.
This demonstrates clearly that the scalarized cqOS-black holes obtained from GB$^-$ scalarization are linearly stable against scalar perturbations, completing a consistent picture of their existence, thermodynamic behavior, and dynamical stability.
This is compared to the case that the fundamental branch of scalarized black holes obtained from  GB$^+$ scalarization is unstable against the radial perturbation for a quadratic coupling of $ \phi^2$ in the EGBS theory~\cite{Blazquez-Salcedo:2018jnn}.

\section{Acknowledgments}

W.K. and Y.S.M. are supported by the National Research Foundation of Korea (NRF) grant
 funded by the Korea government (MSIT) (RS-2022-NR069013). H.G. is supported by the Institute for Basic Science (Grant No. IBS-R018-Y1).\\

\appendix
\section{Equations of radial perturbation}
\label{appendix-A}

In the text, we have shown the detail of the radial perturbation.
Here we list the perturbed equations for the scalar equation~(\ref{s-equa}) and Einstein equation~(\ref{equa1}):
\begin{align}
    &\alpha_1 \Phi''-\alpha_1\ddot{\Phi}+\alpha_2\Phi'+\alpha_3\Phi+\alpha_4H_0''+\alpha_5\ddot{H_1}+\alpha_6H_0'+\alpha_7H_1'+\alpha_8H_0+\alpha_9H_1=0,\\
    &\beta_1H_1'+\beta_2\Phi'+\beta_3H_1=0,\\
    &\gamma_1\dot{H_1}+\gamma_2\dot{\Phi}=0,\\
    &\xi_1H_0'+\xi_2\Phi'+\xi_3H_1+\xi_4H_0=0,\\
    &c_1H_0''+c_2H_0'+c_3H_0+c_4\ddot{H_1}+c_5H_1'+c_6H_1+c_7\Phi'=0.
\end{align}
Here, a ``dot'' denotes the derivative with respect to time, while a prime denotes the derivative with respect to the radial coordinate. 
Combining the above relations as well as the background equations of motion, one arrives at Eq.~\eqref{potential}, with the corresponding coefficients given by
\begin{align}
    &\rho_0=\alpha_1c_1,\\
    &\rho_1=(\alpha_2c_1-\alpha_4c_7)-(\alpha_6c_1-\alpha_4c_2)\frac{\xi_2}{\xi_1}-(\alpha_7c_1-\alpha_4c_5)\frac{\beta_2}{\beta_1},\\
    &\rho_2=\alpha_3c_1-(\alpha_9c_1-\alpha_4c_6)\frac{\gamma_2}{\gamma_1}+(\alpha_6c_1-\alpha_4c_2)\frac{\gamma_2\xi_3}{\gamma_1\xi_1}+(\alpha_7c_1-\alpha_4c_5)\frac{\gamma_2\beta_3}{\gamma_1\beta_1}.
\end{align}


\newpage


\begin{thebibliography}{99}

\bibitem{Carter:1971zc}
B.~Carter,
Phys. Rev. Lett. \textbf{26} (1971), 331-333
doi:10.1103/PhysRevLett.26.331

\bibitem{Ruffini:1971bza}
R.~Ruffini and J.~A.~Wheeler,
Phys. Today \textbf{24} (1971) no.1, 30
doi:10.1063/1.3022513

\bibitem{Herdeiro:2015waa}
C.~A.~R.~Herdeiro and E.~Radu,
Int. J. Mod. Phys. D \textbf{24} (2015) no.09, 1542014
doi:10.1142/S0218271815420146
[arXiv:1504.08209 [gr-qc]].



\bibitem{Doneva:2017bvd}
D.~D.~Doneva and S.~S.~Yazadjiev,
Phys. Rev. Lett. \textbf{120} (2018) no.13, 131103
doi:10.1103/PhysRevLett.120.131103
[arXiv:1711.01187 [gr-qc]].


\bibitem{Silva:2017uqg}
H.~O.~Silva, J.~Sakstein, L.~Gualtieri, T.~P.~Sotiriou and E.~Berti,
Phys. Rev. Lett. \textbf{120} (2018) no.13, 131104
doi:10.1103/PhysRevLett.120.131104
[arXiv:1711.02080 [gr-qc]].

\bibitem{Antoniou:2017acq}
G.~Antoniou, A.~Bakopoulos and P.~Kanti,
Phys. Rev. Lett. \textbf{120} (2018) no.13, 131102
doi:10.1103/PhysRevLett.120.131102
[arXiv:1711.03390 [hep-th]].

\bibitem{Herdeiro:2018wub}
C.~A.~R.~Herdeiro, E.~Radu, N.~Sanchis-Gual and J.~A.~Font,
Phys. Rev. Lett. \textbf{121} (2018) no.10, 101102
doi:10.1103/PhysRevLett.121.101102
[arXiv:1806.05190 [gr-qc]].

\bibitem{Myung:2018vug}
Y.~S.~Myung and D.~C.~Zou,
Eur. Phys. J. C \textbf{79} (2019) no.3, 273
doi:10.1140/epjc/s10052-019-6792-6
[arXiv:1808.02609 [gr-qc]].

\bibitem{Myung:2018jvi}
Y.~S.~Myung and D.~C.~Zou,
Phys. Lett. B \textbf{790} (2019), 400-407
doi:10.1016/j.physletb.2019.01.046
[arXiv:1812.03604 [gr-qc]].

\bibitem{Cunha:2019dwb}
P.~Cunha, V.P., C.~A.~R.~Herdeiro and E.~Radu,
Phys. Rev. Lett. \textbf{123} (2019) no.1, 011101
doi:10.1103/PhysRevLett.123.011101
[arXiv:1904.09997 [gr-qc]].


\bibitem{Collodel:2019kkx}
L.~G.~Collodel, B.~Kleihaus, J.~Kunz and E.~Berti,
Class. Quant. Grav. \textbf{37} (2020) no.7, 075018
doi:10.1088/1361-6382/ab74f9
[arXiv:1912.05382 [gr-qc]].


\bibitem{Dima:2020yac}
A.~Dima, E.~Barausse, N.~Franchini and T.~P.~Sotiriou,
Phys. Rev. Lett. \textbf{125} (2020) no.23, 231101
doi:10.1103/PhysRevLett.125.231101
[arXiv:2006.03095 [gr-qc]].


\bibitem{Hod:2020jjy}
S.~Hod,
Phys. Rev. D \textbf{102} (2020) no.8, 084060
doi:10.1103/PhysRevD.102.084060
[arXiv:2006.09399 [gr-qc]].

\bibitem{Zhang:2020pko}
S.~J.~Zhang, B.~Wang, A.~Wang and J.~F.~Saavedra,
Phys. Rev. D \textbf{102} (2020) no.12, 124056
doi:10.1103/PhysRevD.102.124056
[arXiv:2010.05092 [gr-qc]].


\bibitem{Doneva:2020nbb}
D.~D.~Doneva, L.~G.~Collodel, C.~J.~Kr\"uger and S.~S.~Yazadjiev,
Phys. Rev. D \textbf{102} (2020) no.10, 104027
doi:10.1103/PhysRevD.102.104027
[arXiv:2008.07391 [gr-qc]].

\bibitem{Lai:2022spn}
M.~Y.~Lai, Y.~S.~Myung, R.~H.~Yue and D.~C.~Zou,
Phys. Rev. D \textbf{106} (2022) no.4, 044045
doi:10.1103/PhysRevD.106.044045
[arXiv:2206.11587 [gr-qc]].

\bibitem{Lai:2022ppn}
M.~Y.~Lai, Y.~S.~Myung, R.~H.~Yue and D.~C.~Zou,
Phys. Rev. D \textbf{106} (2022) no.8, 084043
doi:10.1103/PhysRevD.106.084043
[arXiv:2208.11849 [gr-qc]].

\bibitem{Belkhadria:2025lev}
Z.~Belkhadria and S.~Mignemi,
Phys. Rev. D \textbf{112} (2025) no.4, 044015
doi:10.1103/ln23-nhpn
[arXiv:2506.12137 [gr-qc]].

\bibitem{Lewandowski:2022zce}
J.~Lewandowski, Y.~Ma, J.~Yang and C.~Zhang,
Phys. Rev. Lett. \textbf{130} (2023) no.10, 101501
doi:10.1103/PhysRevLett.130.101501
[arXiv:2210.02253 [gr-qc]].

\bibitem{Piechocki:2020bfo}
W.~Piechocki and T.~Schmitz,
Phys. Rev. D \textbf{102} (2020) no.4, 046004
doi:10.1103/PhysRevD.102.046004
[arXiv:2004.02939 [gr-qc]].

\bibitem{Shi:2024vki}
Z.~Shi, X.~Zhang and Y.~Ma,
Phys. Rev. D \textbf{110} (2024) no.10, 104074
doi:10.1103/PhysRevD.110.104074
[arXiv:2408.15821 [gr-qc]].

\bibitem{Dong:2024hod}
S.~H.~Dong, F.~Hosseinifar, F.~Studni{\v{c}}ka and H.~Hassanabadi,
Phys. Lett. B \textbf{860} (2025), 139182
doi:10.1016/j.physletb.2024.139182

\bibitem{Gong:2023ghh}
H.~Gong, S.~Li, D.~Zhang, G.~Fu and J.~P.~Wu,
Phys. Rev. D \textbf{110} (2024) no.4, 044040
doi:10.1103/PhysRevD.110.044040
[arXiv:2312.17639 [gr-qc]].

\bibitem{Zinhailo:2024kbq}
A.~F.~Zinhailo,
Fortsch. Phys. \textbf{73} (2025) no.11, e70038
doi:10.1002/prop.70038

\bibitem{Ou:2025bbv}
M.~Ou and X.~Zhang,
Phys. Rev. D \textbf{112} (2025) no.12, 126016
doi:10.1103/d4fp-nwbf
[arXiv:2508.01183 [gr-qc]].

\bibitem{Ye:2023qks}
J.~P.~Ye, Z.~Q.~He, A.~X.~Zhou, Z.~Y.~Huang and J.~H.~Huang,
Phys. Lett. B \textbf{851} (2024), 138566
doi:10.1016/j.physletb.2024.138566
[arXiv:2312.17724 [gr-qc]].

\bibitem{Luo:2024nul}
S.~Luo and C.~Li,
Phys. Rev. D \textbf{110} (2024) no.12, 124042
doi:10.1103/PhysRevD.110.124042
[arXiv:2409.16323 [gr-qc]].

\bibitem{Yang:2025esa}
S.~Yang, Y.~P.~Zhang, L.~Zhao and Y.~X.~Liu,
Eur. Phys. J. C \textbf{86} (2026) no.1, 35
doi:10.1140/epjc/s10052-026-15284-0
[arXiv:2509.24835 [gr-qc]].

\bibitem{Chen:2025wze}
L.~Chen and S.~Jiang,
Phys. Lett. B \textbf{866} (2025), 139522
doi:10.1016/j.physletb.2025.139522

\bibitem{Myung:2025pmx}
Y.~S.~Myung,
Eur. Phys. J. C \textbf{85} (2025) no.7, 745
doi:10.1140/epjc/s10052-025-14472-8
[arXiv:2505.19450 [gr-qc]].



\bibitem{Mazharimousavi:2025lld}
S.~H.~Mazharimousavi,
Eur. Phys. J. C \textbf{85} (2025) no.6, 667
doi:10.1140/epjc/s10052-025-14410-8
[arXiv:2502.10457 [gr-qc]].


\bibitem{Hassaine:2008pw}
M.~Hassaine and C.~Martinez,
Class. Quant. Grav. \textbf{25} (2008), 195023
doi:10.1088/0264-9381/25/19/195023
[arXiv:0803.2946 [hep-th]].





\bibitem{Dotti:2004sh}
G.~Dotti and R.~J.~Gleiser,
Class. Quant. Grav. \textbf{22} (2005), L1
doi:10.1088/0264-9381/22/1/L01
[arXiv:gr-qc/0409005 [gr-qc]].



\bibitem{Doneva:2018rou}
D.~D.~Doneva, S.~Kiorpelidi, P.~G.~Nedkova, E.~Papantonopoulos and S.~S.~Yazadjiev,
Phys. Rev. D \textbf{98} (2018) no.10, 104056
doi:10.1103/PhysRevD.98.104056
[arXiv:1809.00844 [gr-qc]].

\bibitem{Myung:2018iyq}
Y.~S.~Myung and D.~C.~Zou,
Phys. Rev. D \textbf{98} (2018) no.2, 024030
doi:10.1103/PhysRevD.98.024030
[arXiv:1805.05023 [gr-qc]].

\bibitem{Brihaye:2019kvj}
Y.~Brihaye and B.~Hartmann,
Phys. Lett. B \textbf{792}, 244-250 (2019)
doi:10.1016/j.physletb.2019.03.043
[arXiv:1902.05760 [gr-qc]].

\bibitem{Guo:2025xwh}
H.~Guo, W.~L.~Qian and B.~Wang,
Phys. Rev. D \textbf{112} (2025) no.8, 8
doi:10.1103/861k-193h
[arXiv:2506.02445 [gr-qc]].

\bibitem{Guo:2025flg}
H.~Guo, H.~Liu and Y.~S.~Myung,
Eur. Phys. J. C \textbf{86} (2026) no.2, 204
doi:10.1140/epjc/s10052-026-15407-7
[arXiv:2512.22433 [gr-qc]].

\bibitem{Blazquez-Salcedo:2018jnn}
J.~L.~Bl{\'a}zquez-Salcedo, D.~D.~Doneva, J.~Kunz and S.~S.~Yazadjiev,
Phys. Rev. D \textbf{98}, no.8, 084011 (2018)
doi:10.1103/PhysRevD.98.084011
[arXiv:1805.05755 [gr-qc]].

\bibitem{Minamitsuji:2018xde}
M.~Minamitsuji and T.~Ikeda,
Phys. Rev. D \textbf{99} (2019) no.4, 044017
doi:10.1103/PhysRevD.99.044017
[arXiv:1812.03551 [gr-qc]].


\bibitem{Jansen:2017oag}
A.~Jansen,
Eur. Phys. J. Plus \textbf{132} (2017) no.12, 546
doi:10.1140/epjp/i2017-11825-9
[arXiv:1709.09178 [gr-qc]].



\end{thebibliography}
\end{document}